\documentclass[sigconf]{acmart}
\usepackage{algorithm}
\usepackage{algpseudocode}

\usepackage{enumitem}

\usepackage{multirow}
\usepackage{makecell}
\usepackage{balance}

\AtBeginDocument{%
  \providecommand\BibTeX{{%
    \normalfont B\kern-0.5em{\scshape i\kern-0.25em b}\kern-0.8em\TeX}}}

\copyrightyear{2024}
\acmYear{2024}
\setcopyright{acmlicensed}
\acmConference[SIGIR '24] {Proceedings of the 47th International ACM SIGIR Conference on Research and Development in Information Retrieval}{July 14--18, 2024}{Washington, DC, USA.}
\acmBooktitle{Proceedings of the 47th International ACM SIGIR Conference on Research and Development in Information Retrieval (SIGIR '24), July 14--18, 2024, Washington, DC, USA}
\acmISBN{979-8-4007-0431-4/24/07}
\acmDOI{10.1145/3626772.3661348}

\begin{document}

\title{ECAT: A Entire space Continual and Adaptive Transfer Learning Framework for Cross-Domain Recommendation}


\author{Chaoqun Hou}
\authornote{Corresponding author.}
\email{hcq.hcq@taobao.com}
\authornotemark[2]
\affiliation{%
  \institution{Alibaba Group}
  % \streetaddress{P.O. Box 1212}
  \city{Hangzhou}
  \country{China}
  \postcode{43017-6221}
}

\author{Yuanhang Zhou}
\authornote{Both authors contributed equally to this research.}
\email{zhouyuanhang.zyh@taobao.com}
\affiliation{%
  \institution{Alibaba Group}
  % \streetaddress{1 Th{\o}rv{\"a}ld Circle}
  \city{Hangzhou}
  \country{China}
}

\author{Yi Cao}
\email{dylan.cy@taobao.com}
\affiliation{%
  \institution{Alibaba Group}
  % \streetaddress{1 Th{\o}rv{\"a}ld Circle}
  \city{Hangzhou}
  \country{China}
}

\author{Tong Liu}
\email{yingmu@taobao.com}
\affiliation{%
  \institution{Alibaba Group}
  % \streetaddress{1 Th{\o}rv{\"a}ld Circle}
  \city{Hangzhou}
  \country{China}
}

% \author{xxx}
% \authornote{Corresponding author.}
% \email{xx.xx@xxx.xxx}
% \affiliation{%
%   \institution{xxx xxx}
%   % \streetaddress{1 Th{\o}rv{\"a}ld Circle}
%   \city{xxxx}
%   \country{xxxx}
% }

% \author{xxx}
% \authornote{Corresponding author.}
% \email{xx.xx@xxx.xxx}
% \affiliation{%
%   \institution{xxx xxx}
%   % \streetaddress{1 Th{\o}rv{\"a}ld Circle}
%   \city{xxxx}
%   \country{xxxx}
% }

% \author{xxx}
% \email{xx.xx@xxx.xxx}
% \affiliation{%
%   \institution{xxx xxx}
%   % \streetaddress{1 Th{\o}rv{\"a}ld Circle}
%   \city{xxxx}
%   \country{xxxx}
% }

% \author{xxx}
% \email{xx.xx@xxx.xxx}
% \affiliation{%
%   \institution{xxx xxx}
%   % \streetaddress{1 Th{\o}rv{\"a}ld Circle}
%   \city{xxxx}
%   \country{xxxx}
% }

% \renewcommand{\shortauthors}{Hou and Zhou, et al.}
% \renewcommand{\shortauthors}{xx and xx, et al.}
\renewcommand{\shortauthors}{Chaoqun Hou, Yuanhang Zhou, Yi Cao, \& Tong Liu}

\begin{abstract}
  In industrial recommendation systems, there are several mini-apps designed to meet the diverse interests and needs of users. The sample space of them is merely a small subset of the entire space, making it challenging to train an efficient model. In recent years, there have been many excellent studies related to cross-domain recommendation aimed at mitigating the problem of data sparsity. However, few of them have simultaneously considered the adaptability of both sample and representation continual transfer setting to the target task. To overcome the above issue, we propose a \textbf{E}ntire space \textbf{C}ontinual and \textbf{A}daptive \textbf{T}ransfer learning framework called ECAT which includes two core components: First, as for sample transfer, we propose a two-stage method that realizes a coarse-to-fine process. Specifically, we perform an initial selection through a graph-guided method, followed by a fine-grained selection using domain adaptation method. Second, we propose an adaptive knowledge distillation method for continually transferring the representations from a model that is well-trained on the entire space dataset. ECAT enables full utilization of the entire space samples and representations under the supervision of the target task, while avoiding negative migration. Comprehensive experiments on real-world industrial datasets from Taobao show that ECAT advances state-of-the-art performance on offline metrics, and brings \textcolor{black}{+}13.6\% CVR and \textcolor{black}{+}8.6\% orders for Baiyibutie, a famous mini-app of Taobao.
\end{abstract}





% \ccsdesc[500]{Information systems → Data mining}




\maketitle

%%%%%%%%%%%%%%%%%%%%%%%%%%%%%%%%%%%%%%%%%%%%%%%%%%%%%%%%
\section{Introduction}
Recommendation systems~(RS) have played a significant role in e-commerce platforms, and their efficiency is closely related to the accuracy of click-through rate~(CTR) prediction. In recent years, thanks to the continuous improvements in computational power and the increasing volume of datasets, numerous outstanding single-domain CTR models~\cite{cheng2016wide,guo2017deepfm,chen2022efficient,pi2020search,zhang2021deep} have achieved impressive results. At large e-commercial companies, there are several mini-apps designed to meet the diverse interests and needs of users. However, these mini-apps all encounter a common issue: the target domain has relatively  sparse samples, making it challenging to train the complex CTR model, especially the representations of ID categorical features~(i.e., item ID and user ID). Take Taobao for instance, Baiyibutie is a mini-app that contributes billions of daily page views by exclusively selling brand-discounted products. The sample size of Baiyibutie is less than 1\% of the entire Taobao domain. Therefore, exploring how cross-domain transfer learning can utilize the abundant information available in data-rich domains to enhance the data-sparse domains has emerged as an important research focus in the industry. The traditional cross-domain~\cite{li2023one,huan2023samd,chen2023knowledge,yang2023gradient,mu2023hybrid,tian2023multi,zhao2023m5,gao2023autotransfer} recommendation can be categorized into two paradigms: sample transfer and parameter transfer from the well-trained source model.

\textbf{In the sample transfer paradigm}, multi-task learning methods ~\cite{zhang2021survey,xie2022multi,ma2018modeling,tang2020progressive,sheng2021one,hu2018conet,ouyang2020minet,zou2022automatic} are typically employed to enhance performance across all domains by combining the source and the target samples. However, despite the fact that this paradigm has achieved commendable results in many scenarios, it still has some evident limitations in certain situations. For instance, in scenarios where the sample size of source domain is hundreds of times larger than that of the target domain, the training process can be easily dominated by the source domain, resulting in insufficient training in the target domain. Another issue is that introducing the source domain samples of such a large scale could significantly increase complexity. Therefore, the core objective should be to enhance the performance of the target task by selecting samples that are deemed valuable. 

\textbf{In the parameter transfer paradigm}, pre-training \& fine-tuning methods~\cite{hu2019multi,hu2020discriminative,chen2021user} are more efficient and effective. Specifically, the initialization parameters of the target model are obtained by loading a pre-trained source model, followed by fine-tuning with samples from the target domain. However, an evident issue is that merely fine-tuning with sparse samples from the target domain can easily lead the target model to settle into a sub-optimal local minimum. Therefore, it's crucial to measure the value of the source model's parameters for the target task. Another issue is that few studies considering the setting of Continual Transfer Learning~(CTL)~\cite{wang2020continuously,de2021continual,rusu2016progressive,liu2023continual}, resulting in an inability to continuously utilize the newest information of the source model. CTNet~\cite{liu2023continual} accomplishes continuous transfer by treating the latest source domain representations through an adapter layer. However, in most real-world RSs, user behavioral sequences hold great potential in boosting the performance of the CTR model. CTNet ignores the representations of user behavior sequences in the source model. Furthermore, the target model outperforms the source model on certain samples. Therefore, we need the source model to provide valuable incremental information for these samples that the target model cannot handle well. For these samples that the target model can predict more accurately, we should minimize intervention.

To better solve the above issues in cross-domain modeling, we propose a \textbf{E}ntire space \textbf{C}ontinual and \textbf{A}daptive \textbf{T}ransfer learning framework (\textbf{ECAT}). As shown in Figure~\ref{fig:my_label}, the ECAT framework mainly includes two parts: sample transfer and representations continual transfer. Specifically, we perform a coarse sample selection through a graph guided method, followed by a fine-grained selection using domain adaptation method. During the training process, we continuously transfer valuable information from the source model using an adaptive knowledge distillation method. During the online inference process, the only additional component introduced is the adapter layers, which have a very small complexity. 

To summarize, the main contributions include: 

\begin{itemize}[leftmargin=*]
    \item We propose an ECAT framework, which enables full utilization of the source domain samples and representations under the supervision of the target task, while alleviating negative migration.
    
    \item We propose a two-stage method that realizes a coarse-to-fine process for sample transfer~(GST\,\&\,DA), which enables ECAT to efficiently select samples that are valuable for the target task.
    
    \item We propose an adaptive knowledge distillation method~(AKD-CT) for continually transferring the representations from a source model that is well-trained on the entire space dataset, which allows the ECAT framework to adaptively decide whether to incorporate representational information from the source model.
    
    \item We evaluate ECAT on the Taobao industrial dataset. Comprehensive experiments show that ECAT advances state-of-the-art performance on offline metrics, and brings +13.6\% CVR and +8.6\% orders for Baiyibutie, a mini-app of Taobao.
\end{itemize}

%%%%%%%%%%%%%%%%%%%%%%%%%%%%%%%%%%%%%%%%%%%%%%%%%%%%%%%%
\section{Methods}
\subsection{Problem Definition}
Mathematically, we represent samples from the source domain and the target domain as $D_s=(x_i^s, y_i^s)$ and $D_t=(x_i^t, y_i^t)$ respectively, where $x^s \in R^{d_s}$ and  $x^t \in R^{d_t}$. Label $y_i^s$ and $y_i^t \in \{0, 1\}$ indicate whether the $item_i$ was purchased or not. It is worth mentioning that we have established the capability to acquire samples from the entire domain of Taobao. $S$ is a continually well-trained model on $D_s$, capable of learning new distributions in a timely manner. In this study, our goal is to train a model $T$ using $D_t$ while considering the incremental information in areas including sample transfer from $D_s$ and representation transfer from $S$. Furthermore, we represent $D_s$ through a graph $G_s=(V_s, E_s)$, where $V_s=\{ u_1^s, i_1^s, ...,u_n^s, i_n^s \}$ denotes the user and item node in the graph of source domain. Edge $e_{ij} \in E_s$ denotes that $user_i$ has clicked or purchased on $item_j$. In other words, we can identify the corresponding samples $D_s$ through the nodes and edges of $G_s$. Similarly, we define $G_t=(V_t, E_t)$ according to $D_t$.

% Mathematically, we represent the sample from source domain and target domain as $D_s =(x_i^s, y_i^s)$ and $D_t =(x_i^t, y_i^t)$ respectively, where $x_s \in R^{d_s}$ and $x_t \in R^{d_t}$. $y_i^s \in \{0, 1\}$ and $y_i^t \in \{0, 1\}$corresponds to the label, indicating whether the$item_i$was purchased or not from the source domain and the target domain. It is worth mentioning that we have established a capability to acquire samples from the entire domain of Taobao. $S$ is a continually well-trained model on$D_s$, capable of learning new distributions in a timely manner. In this study, our goal is to train a model $T$using$D_t$ while considering the incremental information in areas including sample transfer from$D_s$and representation transfer from $S$. Furthermore, we represent $D_s$ through a graph network $G_s =(V_s, E_s)$, where $V_s = \{ u_1^s, u_2^s, ...,u_n^s, i_n^s \}$denotes the user and item node in the graph of source domain. $e_{ij} \in E_s$denotes that $user_i$ has clicked or purchased on $item_j$. In other words, we can identify the corresponding samples$D_s$through the nodes and edges of $G_s$. Similarly, we can define $G_t =(V_t, E_t)$ according to $D_t$. 

\begin{figure*}
  \centering
  \includegraphics[width=\linewidth]{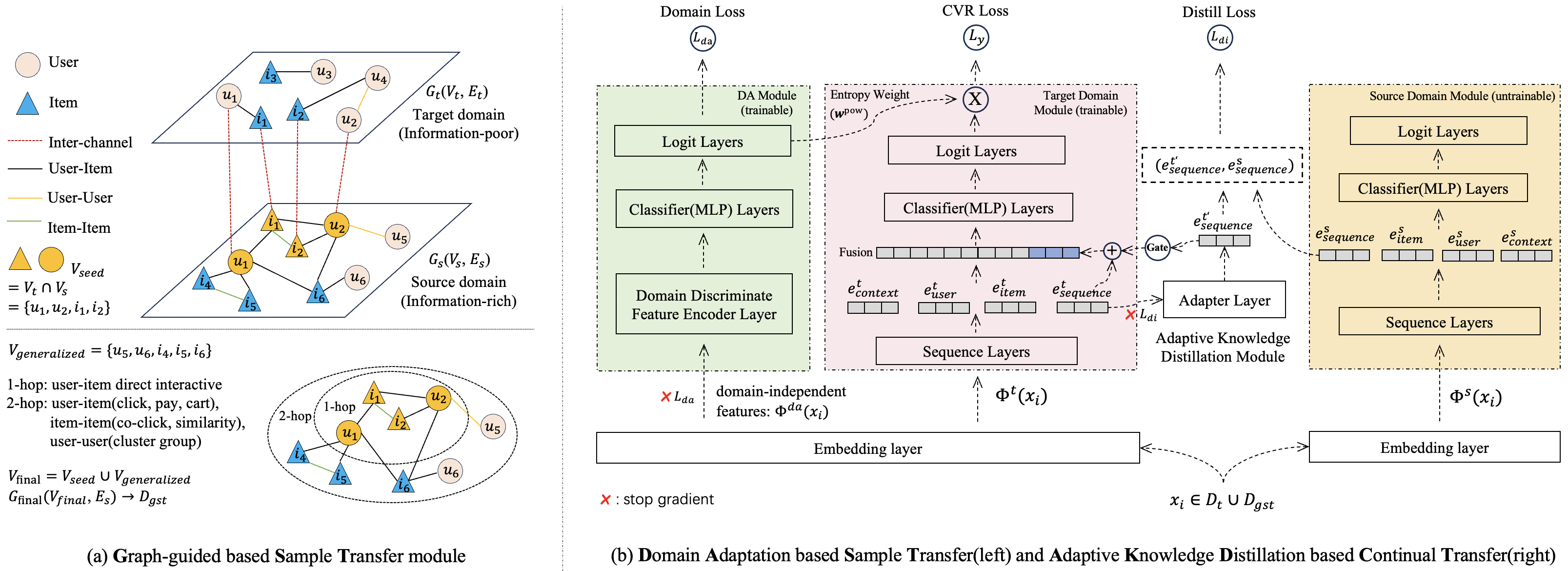}
  \caption{The illustration of ECAT~(Entire space Continual and Adaptive Transfer) framework. ECAT is composed of three parts: First, the Graph-guided module~(a) and the Domain Adaption~(b-left) are aimed to transfer samples. Second, the target model is trained daily. Third, the Adaptive Knowledge Distillation~(b-right) module is for transferring representations continually.}
  \Description{}
  \label{fig:my_label}
\end{figure*}

\vspace{-0.2cm}
\subsection{Model Overview}
We have decomposed the ECAT framework process into two serial stages. Initially, figure~\ref{fig:my_label}(a) shows a simple yet effective method called \textbf{G}raph-guided based \textbf{S}ample \textbf{T}ransfer~(GST), which aims to select samples from $D_s$ with a similar distribution to $D_t$. 
The GST can incorporate sample relevance by leveraging prior heuristic insights or measure it through representational learning via graph neural networks. In this paper, we focus on the area of e-commerce recommendation, there inherently exists plenty of valuable prior knowledge. For example, a direct browse, click or purchase of an item by a user acts as a one-hop link, while a two-hop link can be established between two items through a co-click relationship by users. Moreover, GST is versatile and capable of employing suitable strategies based on the specific domain, or even training a graph representation network model. 
Subsequently, from left to right in figure~\ref{fig:my_label}(b), the diagram sequentially illustrates the Domain Adaption~(DA) module for assessment of incremental value that samples from $D_{gst}$ contribute to $T$, and 
the Adaptive Knowledge Distillation~(AKD-CT) module is designed to assess the incremental value that representations from the well-trained source model $S$. 
More detailed exposition will be delineated in the subsequent discourse.

\vspace{-0.2cm}
\subsection{Graph guided and Domain Adaptation based Sample Transfer}

% In this section, we will introduce a two-stage method that realizes a coarse-to-fine sample transfer process. 

\textbf{Graph guided Module}: Incorporating the findings of many related studies~\cite{crawshaw2020multi,xie2022multi}, the evidence suggests that a model trained using samples from the combined source and target domains, typically results in insufficient training to the target domain. This phenomenon is particularly pronounced in the context of this study, where the sample size of $D_s$ is hundreds of times greater than $D_t$. Considering that, in our scenario, the target domain is a sub-channel of the source domain~($V_t \subset V_s$), the heuristic GST approach is markedly effective and appropriate for our domain-specific challenges. Specifically, as shown in figure~\ref{fig:my_label}(a), we initiate the process by directly mapping $V_{seed}=V_t \cap V_s$ onto $G_s$, anchoring the alignment through the correspondence of the same IDs between the domains. Subsequently, within $G_s=(V_s, E_s)$, 
we expand to include more nodes $V_{generalized}$, which are similar to the target domain, by exploiting one-hop~(i.e., click or pay relationships) and two-hop~(i.e., co-click or group cluster) connectivity.
Finally, within the context of e-commerce recommendation systems, the relevant sample can be identified by specifying a distinct $user_i$ and $item_j$. In other words, we can convert $G^{'}_s = (V_{seed} \cup V_{generalized}, E_s)$  to $D_{gst}$.

\textbf{Target Domain Module}: 
The structure of the target model $T$ is similar to ETA~\cite{chen2022efficient}, which includes four parts. 
First, the embedding layer maps features to representations of a specific dimension, primarily including categorical features and numerical features. 
It is worth mentioning that the categorical features are extremely important and require a substantial number of samples for effective training. 
It is the significant reason why we introduce the well-trained model $S$ across the entire space. Subsequently, long and short-term user behavioral sequences are mapped into higher semantic representations through the sequence layers. 
Finally, we can get the score after successively passing through the classification and logit layers. 
$L_{y}$ is usually a binary cross-entropy loss function. 
\begin{align}
L_{y} = \frac{1}{n_{t}+n_{gst}} \sum_{x_i^{t} \in D_{t} \cup D_{gst}} L_{ce} \left[G^t(\Phi^t(x_i^{t})), y_i^{t}\right],
\end{align}
where $n_{t}$ and $n_{gst}$ are the sample size of $D_t$ and $D_{gst}$ respectively. 
$G^t$ denotes the output of samples pass sequentially from the representation to the logit layers of $T$. $\Phi^t$ is designed to map samples from different feature dimensions to the same feature space, such as attention maps~\cite{komodakis2017paying}.

\textbf{Domain Adaptation Module}: To select samples from $D_{gst}$ that better fit the distribution of $D_{t}$, as shown in figure~\ref{fig:my_label}(b), we refine the sample selection by incorporating a Domain Adaption module. The training dataset is $D_{da}=D_t \cup D_{gst}=(x_i^{da}, y_i^{da})$ and $\Phi(x_i^{da})$ denotes domain-independent features. Label $y_{i}^{da} \in \{0, 1\}$ indicates whether the sample $x_i^{da}$ belongs to $D_{t}$ or $D_{gst}$. The optimization objective $L_{da}$ is a binary cross-entropy loss function. 

The DA module is effective due to three key factors: First, to avoid feature bias, we ensure the effectiveness of the discriminator by solely using domain-independent features. Second, to avoid model bias towards the source domain, we select samples similar to the target domain distribution through GST. Third, to prevent the target model from being influenced by irrelevant gradients, we stop the gradients produced by the DA on the target model. 

Up to this point in our discussion, we have been able to achieve satisfactory results in sample transfer. However, as time progresses, target model $T$ will gradually forget the representations obtained through one-time warm up from $S$, while the representation of $S$ also continues to update. 
We will solve this issue in the next section.

\vspace{-0.25cm}
\subsection{Adaptive Knowledge Distillation based Continual representation Transfer}

\textbf{Source Domain Module}: To provide incremental information, we introduce a source model $S$ that has been well-trained in the entire space. During the training process of $T$, $S$ only executes forward propagation, which entails a low computational complexity. As shown in Figure~\ref{fig:my_label}(b-right), $S$ and $T$ have identical architecture. 

\textbf{AKD-CT Module}: Inspired by CTNet, ECAT endeavors to enhance the target model performance through CTL setting. 
ECAT differs in that it further transfers all layers from the embedding layers to the logit layers, particularly sequence layer representations.
To achieve this, we propose an Adaptive Knowledge Distillation based Continual Transfer (AKD-CT) method. Figure~\ref{fig:my_label}(b-right) illustrates the training process of AKD-CT that showcases the representations distillation of the sequence layers. 
Specifically, we obtain the representations of various behavioral sequence features after passing through the embedding and sequence layers. 
$e_{seq}^t$ and $e_{seq}^s$ denote the sequence representations obtained from $T$ and $S$, respectively. 
Subsequently, $e_{seq}^t$ is the input of adapter layers to obtain $e_{seq}^{t'}$, which then distill knowledge from $S$ under the supervision of $L_{di}$. We use cosine similarity loss to pull $e_{seq}^{t'}$ and $e_{seq}^{s}$ more similar. 
To prevent noise from the distillation process, we stop conducting gradient to $T$.
We have obtained the incremental information $e_{seq}^{t'}$, all that remains is appropriately fusing $e_{seq}^{t'}$ into $T$.

\textbf{The Design of Adaptive Knowledge Distillation} is three-fold: First, considering that $T$ may have better discrimination for certain samples than $S$. We introduce an adaptive gate network to assess the value of $e_{seq}^{t'}$ for $T$. Specifically, we concatenate $e_{seq}^{t'}$, $e_{seq}^t$ and the entropy from $T$ as the input of the adaptive gate network to generate fusion weight. With the supervision of loss $L_{y}$, the fusion weight could indicate the importance of $e_{seq}^{t'}$. Second, our objective is not to distill representations from the source model that are merely similar, but rather those are more suitably adapted to $T$. Therefore, each sample is associated with a distillation intensity $w_i^{pow}$ that governs the degree to which $e_{seq}^{t'}$ approximates $e_{seq}^{s}$. Ideally, the distillation intensity would be higher for samples that $T$ finds hard to predict. After numerous experiments, we adopt the cos similarity to calculate $w_i^{pow}$. Third, our primary task is to enhance the performance of $T$. Therefore, the adapter layers responsible for generating $e_{seq}^{t'}$ are also subject to supervision from $L_y$.

In summary, the final loss of ECAT is as follows:
\begin{align}
L_{ECAT}=\textbf{w}^{da}*L_y+\alpha*\textbf{w}^{pow}*L_{di}+\beta*L_{da},
\end{align}
where $\alpha and \beta$ are hyperparameter that controls the weight of corresponding loss, $\textbf{w}^{da}$ is the entropy value from the DA module for each sample and $\textbf{w}^{pow}$ is the distillation intensity for each sample.

% \begin{align}
% L_{ecat}=W^{da}*L_y+\alpha*W^{pow}*L_{di}+\beta*L_{da},
% \end{align}

%%%%%%%%%%%%%%%%%%%%%%%%%%%%%%%%%%%%%%%%%%%%%%%%%%%%%%%%
\section{Experiments}
\subsection{Experimental Setup}

\subsubsection{Dataset}
In the absence of suitable public benchmarks for evaluating continual cross-domain prediction, we adopt Taobao industrial datasets to comprehensively compare ECAT and baselines. Therefore, we use the Baiyibutie from Taobao as the target domain, which generates millions CVR samples every day, accounting for less than 1\% of the entire space of Taobao. The users and items in the source domain and the target domain partially overlap, but the data distribution is very different. In this study, we utilize target domain samples spanning 90 days, amounting to a total of 120 million samples. Similarly, taking the entire space as the source domain, we have accumulated a total of 66 billion samples. During A/B testing, ECAT serves over hundreds of thousands of users daily.

%Pretrain\,\&\,Finetune~\cite{hu2019multi}
\subsubsection{Baseline Models} We compare the samples transfer methods like simple merge $D_s$ and $D_t$, DANN~\cite{ganin2015unsupervised}, the representations transfer methods including Shared Bottom, MMoE~\cite{ma2018modeling}, PLE~\cite{tang2020progressive}, and the continual learning setting method like CTNet~\cite{liu2023continual}.

\subsubsection{Implementation Details} To ensure the fairness of the experiments, all single domain methods employs the ETA~\cite{chen2022efficient} as the architecture, including the target model and source model. Specifically, we use AdagradDecayV2~\cite{duchi2011adaptive} as the optimizer. Learning rate is set to 0.01 and the batch size is 1024. The dimension of MLP Layers is set to 1024, 512 and 256. Following previous work, we adopt AUC to measure the CVR prediction performance in offline evaluation.

\subsection{Offline Evaluation}

As shown in Table~\ref{main-result}, our ECAT (GST\,\&\,DA and AKD-CT) achieves the best performance among all baselines. More specifically,
% \textcolor{red}{The offline evaluation result is showed in Table~\ref{main-result}, we can see that,}

(1) ECAT achieves the best performance (AUC=0.8348) compared to both single-domain and cross-domain methods, with its core advantage being that ECAT simultaneously considers the adaptability of both sample and representation for the target task.

(2) In terms of sample transfer: ECAT enhances the performance of each method, including CTNet (AUC from 0.8307 to 0.8327), by transferring valuable samples from $D_s$ through GST\,\&\,DA. Besides, directly merging $D_s$ with $D_t$ leads to performance degradation.

(3) In terms of continual representation transfer: ECAT further improves performance through the adaptive capabilities of AKD-CT module, which under the supervision of the target task, continuously transfers valuable representation information from $S$. Even with the same sample transfer strategy, the effectiveness of AKD-CT (AUC=0.8348) surpasses that of CTNet (AUC=0.8327).

% (1) Among all sample transfer methods, GST achieves best performance. As GST selects the high-quality source domain samples that can improve the model performance on target domain.

% (2) AKD-CT performs better than the single-domain methods and cross-domain recommendation methods, demonstrating the effectiveness of continuous transfer for cross-domain recommendation.
% % indicating the effectiveness of the cross-domain recommendation methods.

% (3) Among the continuous transfer methods, AKD-CT performs better than CTNet, illustrating the effectiveness of the adaptive distillation module and the distillation on behavior sequences.

% (4) When both GST and AKD-CT achieved remarkable results, we combined these two methods into our ECAT framework and achieved the best cross-domain recommendation performance.

\vspace{-0.2cm}
\begin{table}[!ht]
    \setlength{\abovecaptionskip}{0.1cm}
    \small
    \caption{Offline results of various methods.}
    \centering
    \begin{tabular}{l|ccc}
    \hline
        \multirow{2}{*}{Method} & \multicolumn{3}{c}{Sample Transfer Setting} \\ \cline{2-4}
        % - & $D_t$ & $D_t \cup D_s$ & $D_t \cup D_{gst}$ \\ \hline
        
         & Only $D_t$ & Merge $D_s$ and $D_t$ & GST\,\&\,DA \\ \hline
        % Method & \multicolumn{3}{c}{AUC} \\ \hline
        % Method & AUC & AUC & AUC \\ \hline
        Target Model & 0.8284 & 0.8276 & 0.8301 \\ 
        DANN & 0.8289 & 0.8287 & 0.8307 \\ 
        Shared Bottom & 0.8297 & 0.8301 & 0.8312 \\ 
        MMOE & 0.8287 & 0.8286 & 0.8303 \\ 
        PLE & 0.8303 & 0.8298 & 0.8324 \\ \hline
        CTNet & 0.8307 & 0.8302 & 0.8327 \\ 
        AKD-CT & 0.8327 & 0.8321 & \textbf{0.8348} \\ \hline
        % AKD-CT$^{all}$ +KF  & 0.8319 & 0.8313 & 0.8341 \\ 
        % AKD-CT$^{all}$ +KF +RSM & 0.8327 & 0.8321 & \textbf{0.8348} \\ \hline
    \end{tabular}
    \label{main-result}
\end{table}

\vspace{-0.3cm}
\subsection{Research Questions}

\subsubsection{RQ1: How to prove the adaptive capability of AKD-CT model?}
\ 
\newline
\indent 
% \textcolor{red}{As shown in Table~\ref{ablation_exp}, \emph{ECAT without gate} gets a worse performance, the reason is that the adaptive gate network can retain knowledge and filter out noise within the incremental information, enhancing the performance of $T$ on highly uncertain samples. Second, \emph{ECAT without intensity} achieves a worse performance, it indicates that the similarity based intensity could adaptively modulate the intensity based on the representations of $T$ and $S$, ensure that knowledge more suitable for $T$ is better distilled. Finally, \emph{ECAT without selective} achieves a performance decrease, the reason is that selective mechanism directly employs the source representation for transfer when the incremental information is inadequately trained, thereby avoiding the introduction of noise during training.}
Table~\ref{ablation_exp} shows that (1)~AKD-CT drops performance without gate. The reason is that the gate network assesses the importance of incremental information for $T$, providing valuable incremental representation information for samples with higher uncertainty and lower confidence. (2)~The absence of distillation intensity in AKD-CT results in poorer result, which suggests that an intensity-based strategy facilitates the distillation of representations more suited to $T$.

% (2)~The absence of intensity mechansim in AKD-CT results in poorer result, 
% confirming that adaptive distillation intensity can achieve better knowledge distillation, thereby improving the performance of the target model.
% (3)~The lack of a selective mechanism leads to a performance decline, showing its importance in leveraging source representation for transfer to prevent noise during early training stages.

% We conduct the ablation experiment based on some variants of our framework, including (1) \emph{ECAT without gate} removes the adaptive gate network and directly adds the incremental information to target model representations; (2) \emph{ECAT without intensity} removes the representation similarity based distillation intensity mechanism and utilize fixed intensity; (3) \emph{ECAT without selective} removes the selective method and always utilize incremental information.
% As shown in Table~\ref{ablation_exp}, we observe that all ablation experiments lead to performance decrease. It demonstrates the effectiveness of each module in our proposed ECAT method.

\vspace{-0.2cm}
\begin{table}[h]
\setlength{\abovecaptionskip}{0.1cm}
\small
\caption{Ablation experiments on adaptive capability.}
\begin{tabular}{l|c}
\hline
Adaptive Setting   & AUC     \\
\hline
AKD-CT  & \textbf{0.8348}  \\
AKD-CT  without gate     & 0.8331 (-0.20\%) \\
AKD-CT  without intensity     & 0.8342 (-0.07\%) \\
% ECAT without selective      & 0.8341 (-0.08\%) \\
\hline
\end{tabular}
\label{ablation_exp}
\end{table}

\vspace{-0.2cm}
\subsubsection{RQ2: How to prove the necessity of CTL setting?}
\ 
\newline 
\indent
We compare the performance between AKD-CT and CTNet under continuous transfer and one-time transfer setting. Table~\ref{transfer_exp} shows that continuous transfer is better than one-time transfer, which illustrates the necessity of CTL. $\Delta$t is 30 days in this study.
% ACK-CT outperforms CTNet, illustrating the effectiveness of distillation-based continuous transfer.
% \indent We compare the performance between ACK-CT and CTNet under two transfer settings(continuous transfer~\footnote{Continuous transfer: during training process, the representations of source model are continuously transferred to target domain model through distillation or fusion.} and one-time transfer~\footnote{One-time transfer: the target domain model is initialized only once by the source domain model at the beginning of training.}). As shown in Table~\ref{transfer_exp}, continuous transfer is always better than one-time transfer, which illustrates the necessity of CTL. ACK-CT outperforms CTNet, illustrating the effectiveness of distillation-based continuous transfer.

\vspace{-0.2cm}
\begin{table}[!ht]
\setlength{\abovecaptionskip}{0.1cm}
    \small
    \caption{The comparisons between different transfer setting.}
    \centering
\begin{tabular}{c|c|ccc}
\hline
Transfer Setting & Method & t  & t+$\Delta$t & t+2$\Delta$t \\
\hline
\multirow{3}{*}{one-time}  & Base (PLE) & 0.8354 & 0.8372   & 0.8324    \\
& CTNet    & 0.8356 & 0.8373   & 0.8325     \\
& AKD-CT      & 0.8366 & 0.8378   & 0.8336     \\ \hline
\multirow{2}{*}{continual}     & CTNet    & 0.8356 & 0.8375   & 0.8327     \\
& AKD-CT      & 0.8366 & 0.8382   & \textbf{0.8348}    \\  \hline
\end{tabular}
\label{transfer_exp}
\end{table}

\vspace{-0.4cm}
\section{CONCLUSIONS}
In this paper, we introduce the ECAT framework for cross-domain prediction, which not only considers the continual transfer of both samples and representations but also the adaptability of incremental information to the target task. Experiments conducted on a large-scale industrial dataset, along with online A/B testing, confirm its effectiveness in real-world applications. It is noteworthy that ECAT has been deployed in the RS of Taobao to serve numerous marketing channels, including Baiyibutie.

% \begin{table}
%   \caption{Frequency of Special Characters}
%   \label{tab:freq}
%   \begin{tabular}{ccl}
%     \toprule
%     Non-English or Math&Frequency&Comments\\
%     \midrule
%     \O & 1 in 1,000& For Swedish names\\
%     $\pi$ & 1 in 5& Common in math\\
%     \$ & 4 in 5 & Used in business\\
%     $\Psi^2_1$ & 1 in 40,000& Unexplained usage\\
%   \bottomrule
% \end{tabular}
% \end{table}

% \begin{table*}
%   \caption{Some Typical Commands}
%   \label{tab:commands}
%   \begin{tabular}{ccl}
%     \toprule
%     Command &A Number & Comments\\
%     \midrule
%     \texttt{{\char'134}author} & 100& Author \\
%     \texttt{{\char'134}table}& 300 & For tables\\
%     \texttt{{\char'134}table*}& 400& For wider tables\\
%     \bottomrule
%   \end{tabular}
% \end{table*}

\bibliographystyle{ACM-Reference-Format}
\balance
\bibliography{references}

% \appendix






\end{document}